\documentstyle[amssymb,rmp,aps]{revtex}

\begin{document}
\title{\ \ \ \ \ \ \ \ \ \ \ \ Some geometrical aspects of Microcanonical
Distribution.}
\author{L.Vel\'{a}zquez $^{1}$, F.Guzm\'{a}n $^{2}$}
\address{$^{1}$Departamento de Fisica, Universidad de Pinar del Rio, Cuba\\
e-mail: luisberis@geo.upr.edu.cu \\
$^{2}$Departamento de Fisica Nuclear, Instituto Superior de Ciencias y\\
Tecnolog\'{i}as Nucleares, Ciudad Habana, Cuba.\\
e-mail: guzman@info.isctn.edu.cu }
\date{\today}
\maketitle

\begin{abstract}
\ \ In the \ present work\ is \ presented \ some considerations\ for\ a
possible\ generalization of the Microcanonical Thermoestatics $\left(
M.Th\right) $ of \ D.H.E. Gross. The same reveals a geometric aspect that
commonly it has been disregarded so far: {\it the local reparametrization
invariance }. This new characteristic lead to the needing of generalizing
the methods of $M.Th$ to be consequent with this property.
\end{abstract}

\section{Introduction}

In the last years we have been witnesses of the \ development \ of a \ new
extension of the \ Statistical \ Mechanics for non extensive systems: the
Microcanonical Thermostatistic of D.H.E. Gross$\left[ 1\right] $. This
theory do not invoke the thermodynamical limit, and in this way, it permits
to extend the Thermodynamics for the study of the small systems. For this,
it appeals to the definition of entropy given by Boltzmann as function of
the conserved extensive variables. The topology of the curvature of entropy
is arising as an order parameter showing all the possible varieties for the
phase transitions. The same is essentially a geometrical formulation,
characteristic that makes it very elegant and transparent.

In \ the present \ work is presented some considerations for \ a possible
generalization for \ this theory:\ the Multimicrocanonical Distribution. The
same is motivated during the study of the self-confined small systems, for
which the macroscopic state should be necessarily described by several
integrals of movement. This is the case of the nuclear systems, the atomic,
molecular clusters and the selfgravitating astrophysical systems$\left[ 2,3%
\right] $.

Let $\left( I,N,a\right) $ the representation of the macroscopic state of
the system,\ where $I\equiv \left\{ I^{1},I^{2}\ldots I^{n}\right\} $\ are
the set of integrals of movement of the distribution, $N$\ \ the particle
number and $a$\ represents an external parameter. Fallowing the general
ideas of the Microcanonical Thermostatistic, the probability phase density
is given by:\bigskip 
\begin{equation}
\omega _{M}\left( X;I,N,a\right) =\frac{1}{\Omega \left( I,N,a\right) }%
\delta \left( I-I_{N}\left( X;a\right) \right)
\end{equation}
generalizing the ordinary microcanonical distribution, where $\Omega $\ is
the state density. The choice of equal-probability hypothesis allows to take
a totally geometrical version for the statistic mechanics without the
introduction of artificial elements in the theory.

Similarly, we assumed the Boltzmann's definition of entropy:

\begin{equation}
S_{B}\left( I,N,a\right) =\ln \left( \Omega \left( I,N,a\right) \delta
I\right)
\end{equation}
where $\delta I$ is a suitable elemental volume. The Boltzmann definition of
entropy does not require to satisfy the concavity and extensity conditions
that ordinary exhibits the information entropy. How we can see, this
definition is essentially geometrical and transparent in its sense.

Easily we can show the following result:

\begin{eqnarray}
\frac{\partial S_{B}\left( I,N,a\right) }{\partial a} &=&\overrightarrow{%
\nabla _{I}}\cdot \overrightarrow{F_{a}}\left( I,N,a\right) +\overrightarrow{%
\beta }\cdot \overrightarrow{F_{a}}\left( I,N,a\right) \\
&=&\frac{1}{\Omega \left( I,N,a\right) }\overrightarrow{\nabla _{I}}\cdot
\left( \Omega \left( I,N,a\right) \overrightarrow{F_{a}}\left( I,N,a\right)
\right)  \nonumber
\end{eqnarray}
where $\overrightarrow{\beta }=\overrightarrow{\nabla _{I}}S_{B}$\ is the
canonical parameter and $\overrightarrow{F_{a}}=\langle -\overrightarrow{%
\frac{\partial I\left( X;a\right) }{\partial a}}\rangle _{M\text{ \ }}$\ ,
the generalized force.

The similar ordinary expression is:

\begin{equation}
\frac{\partial S_{B}\left( E,N,a\right) }{\partial a}=\beta \cdot
f_{a}\left( E,N,a\right)
\end{equation}
where $\beta =\frac{\partial S_{B}\left( E,N,a\right) }{\partial E}=\frac{1}{%
T}$\ the inverse of temperature and\ $\ f_{a}=\langle -\overrightarrow{\frac{%
\partial H\left( X;a\right) }{\partial a}}\rangle _{\text{ \ }}$the
generalized force.

In order to probe the above result, let us to introduce the Generating
Functional of the Multimicrocanonical Distribution, $G\left( k;I,N,a,\right) 
$: 
\begin{equation}
G\left( k;I,N,a,\right) =\int dX\exp \left( ikX\right) \frac{1}{\Omega
\left( I,N,a\right) }\delta \left( I-I_{N}\left( X;a\right) \right)
\end{equation}

The quantities mean values comes give by: 
\[
O\left( I,N,a\right) =\left. O\left( \widehat{p}\right) G\left(
k;I,N,a,\right) \right| _{k=0} 
\]
where $\widehat{p}=-i\frac{\partial }{\partial k}$ . It is easy to show that
this functional is the solution of the operational equation: 
\begin{equation}
I^{s}\left( \widehat{p};a\right) G\left( k;I,N,a,\right) =I^{s}G\left(
k;I,N,a,\right)
\end{equation}
\[
\left. G\left( k;I,N,a,\right) \right| _{k=0}=1 
\]

This functional satisfy the relation: 
\begin{equation}
\frac{\partial }{\partial a}\left[ \Omega \left( I,N,a\right) G\left(
k;I,N,a,\right) \right] =F_{a}\left( \widehat{p};a\right) \cdot \nabla _{I}%
\left[ \Omega \left( I,N,a\right) G\left( k;I,N,a,\right) \right]
\end{equation}
where $F_{a}\left( X;a\right) =-\frac{\partial }{\partial a}I\left(
X;a\right) $ is the generalized force. Let us probe this relation. 
\[
\frac{\partial }{\partial a}\left[ \Omega \left( I,N,a\right) G\left(
k;I,N,a,\right) \right] =\int dX\exp \left( ikX\right) \left[ -\frac{%
\partial }{\partial a}I\left( X;a\right) \right] \cdot \nabla _{I}\delta
\left( I-I_{N}\left( X;a\right) \right) 
\]

Introducing the operator $F_{a}\left( \widehat{p};a\right) $ and extracting
all the operators outside the integral: 
\[
F_{a}\left( \widehat{p};a\right) \cdot \nabla _{I}\left[ \int dX\exp \left(
ikX\right) \delta \left( I-I_{N}\left( X;a\right) \right) \right] 
\]
we arrive to the final result: 
\[
\frac{\partial }{\partial a}\left[ \Omega G\right] =\widehat{F}_{a}\cdot
\nabla _{I}\left[ \Omega G\right] 
\]

From this relation is derived the Generalized Gibb Identities. Acting on
this relationship the operator of any physical magnitude, $O\left( \widehat{p%
}\right) ,$ we find: 
\begin{equation}
\frac{\partial }{\partial a}\left[ \Omega \overline{O}\right] =\nabla
_{I}\cdot \left[ \Omega \overline{OF_{a}}\right]
\end{equation}

In particular, if $O=1:$ 
\begin{equation}
\frac{\partial }{\partial a}\left[ \Omega \right] =\nabla _{I}\cdot \left[
\Omega \overline{F_{a}}\right]
\end{equation}
Therefore: 
\begin{equation}
\frac{\partial }{\partial a}\left[ \ln \Omega \right] =\nabla _{I}\cdot 
\overline{F_{a}}+\nabla _{I}\left[ \ln \Omega \right] \cdot \overline{F_{a}}
\end{equation}

As we can see, here it appears the term of the divergence of the force. If
we assume the Boltzmann definition of entropy given in Eq. (2), considering $%
\delta I$ constant, we arrived finally to: 
\[
\frac{\partial }{\partial a}S_{B}=\nabla _{I}\cdot \overline{F_{a}}+\nabla
_{I}S_{B}\cdot \overline{F_{a}}
\]

We can see that the generalized force and a $\beta $-canonical parameter
become now in vectors. The presence of the divergency of the force is an
effect of the non-extensive nature of the system. Ordinarily this term
vanishes in the thermodynamical limit:

\begin{equation}
\frac{\overrightarrow{\nabla }\cdot \overrightarrow{F_{a}}}{\overrightarrow{%
\beta }\cdot \overrightarrow{F_{a}}}\sim \frac{1}{N}\rightarrow 0
\end{equation}
but, {\it we can not ignore it here}.

As we can see before, in this formulation appears the concepts of scalar
product and divergency of vectors belonging the geometrical approach. It
motived to develop a geometrical formalism for to work with this
distribution.

\bigskip

\section{Geometrical Aspects}

The geometrical aspects of the theory are summed up by the fallowing facts:

\begin{itemize}
\item  Any function of the movement integrals is itself a movement integral.
If we have a complete independent set of functions of the movement integral
of the distribution, then this set of functions is equivalent to the other
set of movement integrals, representing the same macroscopic state of the
system. That is why it is more exactly to speak about the abstract Space of
movement integrals of the multimicrocanonical distribution, $\Im _{N}$.
Therefore, every physical quantity or behavior has to be equally sketched by
any representation of the space of macroscopic state, $\left( \Im
_{N};a\right) $\ .

\item  It is easy to show that the multimicrocanonical distribution is a
reparametrization invariant. Let $\Re _{I}$ and $\Re _{\phi }$ $\left( \
I\equiv \left\{ I^{1},I^{2}\ldots I^{n}\right\} \ \text{, }\ \phi \equiv
\left\{ \phi ^{1}\left( I\right) ,\phi ^{2}\left( I\right) \ldots \phi
^{n}\left( I\right) \right\} \right) $\ \ two representations of $\Im _{N}$.
By the property of the $\delta $-function we have:
\end{itemize}

\begin{equation}
\delta \left[ \phi \left( I\right) -\phi \left( I_{N}\left( X;a\right)
\right) \right] =\left| \frac{\partial \phi }{\partial I}\right| ^{-1}\delta %
\left[ I-I_{N}\left( X;a\right) \right]
\end{equation}

hence:

\begin{equation}
\widetilde{\Omega }\left( \phi ,N,a\right) =\left| \frac{\partial \phi }{%
\partial I}\right| ^{-1}\Omega \left( I,N,a\right)
\end{equation}

and therefore:

\begin{equation}
\omega _{M}\left( X;\phi ,N,a\right) =\omega _{M}\left( X;I,N,a\right)
\equiv \omega _{M}\left( X;\Im _{N},a\right)
\end{equation}

\begin{itemize}
\item  \ The state density allows to define the invariant measure of the
space : 
\begin{equation}
d\mu =\Omega \left( I,N,a\right) d^{n}I
\end{equation}
although this is the most important measure for the space, this is not the
only one, because there are others like them derived from Poincare
invariants, when we project the N-body phase-space to the space\ $\Im _{N}$.

\item  The movement integrals are defined by the commutativity relation with
the Hamiltonian of the system. In the case of closed systems, the
Hamiltonian is the energy of the system, and this is a conserved quantity.
In the multimicrocanonical distribution it represents one of the coordinate
of the point belonging to $\Im _{N}$, in an specific representation. When we
change the coordinate system, the energy lose its identity. Since we can not
fix the commutativity of the energy with the others integrals, it would be
more correct that all movement integrals commute between them in order to
preserve these conditions. Thus, we make consistently the reparametrization
invariance with the commutativity relations. As we see, still from the
classical point of view, the reparametrization invariance suggests that the
set of movement integrals have to be\ simultaneously measured. In the
quantum case, this is an indispensable request for the correct definition of
the statistical distribution. This property is landed to the classical
distribution by the correspondent principle between the Quantum and the
Classic Physic.
\end{itemize}

This reparametrization freedom is very attractive, since we can choose the
representation of the macroscopic state space more adequate to describe the
statistical system. There are many examples in which an adequate selection
of coordinate system helps to simplify the resolution of one problem.

Let us move now to the Boltzmann's definition of entropy. Obviously\ it has
to be an scalar function. Its value can not depends on the coordinate system
that we use to describe the macroscopic state. It is our interest that this
function characterizes the macroscopic state of the system, it have to
allows us to get information about the ordering of it. In this case, $\delta
I$ can not be arbitrary. It has to be a characteristic of the representation
of the space $\Im _{N}.$

The physical meaning of the state density lead that this function have to be
positive, $\Omega \left( I,N,a\right) \geqslant 0$\ . If the point\ $I$\ is
not a permissible value for the integrals of distribution, then $\Omega
\left( I,N,a\right) =0$\ . But, when we perform the calculations we have to
smooth the $\delta $-function. It provokes that the state density may be
analytically continued and therefore vanishes or is negative in that point.
The points on the boundary of accessible region satisfy the condition:

\qquad 
\begin{equation}
\Omega \left( I,N,a\right) =0
\end{equation}

When we change the coordinate system, these points remain satisfying the
same condition:

\begin{equation}
\widetilde{\Omega }\left( \phi ,N,a\right) =\left| \frac{\partial \phi }{%
\partial I}\right| ^{-1}\Omega \left( I,N,a\right) \Rightarrow \widetilde{%
\Omega }\left( \phi ,N,a\right) =0
\end{equation}
so, we can write before as:

\begin{equation}
\Omega \left( \partial \Im _{N},a\right) =0
\end{equation}
where $\partial \Im _{N}$ are the boundaries of accessible regions.

The existence and topology of these regions do not depend on the
representation of the space. We are interested in describing this kind of
geometrical and topological characteristics because all of them represent
general features of the behavior of the system.

\end{document}